\documentclass[english,aps,manuscript,aps,superscriptaddress]{revtex4-1}
\usepackage[T1]{fontenc}
\usepackage[latin9]{inputenc}
\usepackage{babel}
\usepackage{rotfloat}
\usepackage{bm}
\usepackage{amsmath}
\usepackage{amssymb}
\usepackage{graphicx}
\usepackage{esint}
\usepackage[unicode=true,
 bookmarks=true,bookmarksnumbered=false,bookmarksopen=false,
 breaklinks=false,pdfborder={0 0 0},backref=false,colorlinks=false]
 {hyperref}
\hypersetup{pdftitle={A Monte Carlo Study of Low Energy Electron and Positron Transport in Percus-Yevick and Argon Liquids},
 pdfauthor={W. J. Tattersall, D. G. Cocks, G. Boyle and R. D. White},
 pdfkeywords={Monte Carlo, transport}}

\makeatletter

\providecommand{\tabularnewline}{\\}

\@ifundefined{textcolor}{}
{%
 \definecolor{BLACK}{gray}{0}
 \definecolor{WHITE}{gray}{1}
 \definecolor{RED}{rgb}{1,0,0}
 \definecolor{GREEN}{rgb}{0,1,0}
 \definecolor{BLUE}{rgb}{0,0,1}
 \definecolor{CYAN}{cmyk}{1,0,0,0}
 \definecolor{MAGENTA}{cmyk}{0,1,0,0}
 \definecolor{YELLOW}{cmyk}{0,0,1,0}
}

\makeatletter
\providecommand*{\diff}%
        {\@ifnextchar^{\DIfF}{\DIfF^{}}}
\def\DIfF^#1{%
        \mathop{\mathrm{\mathstrut d}}%
                \nolimits^{#1}\gobblespace
}
\def\gobblespace{%
        \futurelet\diffarg\opspace}
\def\opspace{%
        \let\DiffSpace\!%
        \ifx\diffarg(%
                \let\DiffSpace\relax
        \else
                \ifx\diffarg\[%
                        \let\DiffSpace\relax
                \else
                        \ifx\diffarg\{%
                                \let\DiffSpace\relax
                        \fi\fi\fi\DiffSpace}   
\usepackage{multirow}

\makeatother

\begin{document}

\title{An improved Monte Carlo study of coherent scattering effects of low
energy charged particle transport in Percus-Yevick liquids}

\author{W. J. Tattersall}

\affiliation{Research School of Physics and Engineering, The Australian National
University, Canberra, ACT 0200, Australia}

\affiliation{College of Science, Technology and Engineering, James Cook University,
Townsville 4810, Australia}

\author{D. G. Cocks}

\affiliation{College of Science, Technology and Engineering, James Cook University,
Townsville 4810, Australia}

\author{G. J. Boyle}

\affiliation{College of Science, Technology and Engineering, James Cook University,
Townsville 4810, Australia}

\author{R. D. White}

\affiliation{College of Science, Technology and Engineering, James Cook University,
Townsville 4810, Australia}
\begin{abstract}
We generalize a simple Monte Carlo (MC) model for dilute gases to
consider the transport behavior of positrons and electrons in Percus-Yevick
model liquids under highly non-equilibrium conditions, accounting
rigorously for coherent scattering processes. The procedure extends
an existing technique \nocite{Wojcik2002} {[}Wojcik and Tachiya,
Chem. Phys. Lett. \textbf{363}, 3--4 (1992){]}, using the static structure
factor to account for the altered anisotropy of coherent scattering
in structured material. We identify the effects of the approximation
used in the original method, and develop a modified method that does
not require that approximation. We also present an enhanced MC technique
that has been designed to improve the accuracy and flexibility of
simulations in spatially-varying electric fields. All of the results
are found to be in excellent agreement with an independent multi-term
Boltzmann equation solution, providing benchmarks for future transport
models in liquids and structured systems.
\end{abstract}
\maketitle

\section{Introduction}

The precise behavior of electrons and positrons traveling through
matter is of vital importance in many new and established technologies.
Applications such as solar cells \cite{Brodsky1985}, radiation dosimetry
\cite{Nikjoo1994}, material pore-size classification \cite{Gidley2006}
and positron emission tomography \cite{Buvat2005} all require an
understanding of the fundamental physical processes involved, including
accurate knowledge of energy deposition, macroscopic behaviors, and
loss rates.

Although the behavior of high-energy particles can be simulated quite
accurately with condensed history techniques \cite{Kawrakow1998},
at low energies it is important to model the individual interactions
of particles colliding with the background material, and thereby monitor
discrete energy losses and processes that change the number of particles
in the system. In the systems that we are investigating, the number
density of the charged particles is low enough that the Debye wavelength
greatly exceeds the dimensions of the the system, which is known as
the ``swarm'' limit of an ionized gas \cite{Robson2006}. 

One successful approach to modeling such systems is by solving the
Boltzmann equation \cite{White2011}, which is an equation of continuity
in phase space. Often this approach is limited to idealized geometries,
due to complexities in the numerical solution and application of boundary
conditions. However, many real-world systems are too complex for such
an approach to be effective, and in any case, alternative methods
should ideally be used for verification. 

The pre-eminent alternative is to use Monte Carlo simulations, which
have been widely employed for such purposes \cite{Allen1984,Champion2012,Emfietzoglou2003,Suvakov2008,Munoz2005}
ever since computers have been powerful enough to implement them \cite{Skullerud1968}.
Monte Carlo simulations are very flexible, and can easily include
features from systems that are quite difficult to model in any other
manner, such as interfacial effects, secondary particles, and inhomogenous
media.

Accurate simulations of condensed systems must include the effects
of coherent scattering, where the incoming electrons and positrons
interact with many particles of the system at once. This can occur
when the de Broglie wavelength of the low energy electrons and positrons
is longer than the mean distance between molecules of the condensed
matter \cite{Sakai1988}. It is common to ignore these effects in
Monte Carlo simulations of liquids \cite{Champion2006}, because they
are usually insignificant for electrons and positrons with energies
of greater than $\sim10-20$ eV. However, to accurately treat particle
transport at low energies, we must include these collective effects,
usually by way of the medium's dynamic structure factor $S\left(\Delta\mathbf{k},\Delta\omega\right)$
\cite{Sakai1988}, which contains information about the medium's characteristic
allowed transfers of momentum $\hbar\Delta\mathbf{k}$ and energy
$\hbar\omega$. The dynamic structure is exactly what is measured
by coherent neutron scattering experiments such as \cite{Verkerk1991}. 

The present study will describe a new Monte Carlo implementation that
models structured matter using a static structure factor, $S\left(\Delta\mathbf{k}\right)$,
which is an integrated form of the dynamic structure factor. We first
use a technique described by Wojcik and Tachiya \cite{Wojcik2002}
to incorporate the static structure factor into our Monte Carlo model,
but we assert that this method is only accurate for a certain subset
of cases. We subsequently extend this technique to overcome its limitations.
The Percus-Yevick static structure factor is a simple analytic static
structure factor \cite{Verlet1972} that can be used as a benchmark
to verify the accuracy of our simulation. We have performed simulations
of a number of Percus-Yevick systems at a range of reduced electric
field strengths, and we compare our results with those obtained by
solving the Boltzmann equation detailed in \cite{White2011}. 

We begin this study with a discussion of the Boltzmann equation approach
to coherent scattering. We follow this with a brief description of
typical Monte Carlo collision mechanics for elastic processes in section~\ref{sub:Coherent-elastic-scattering}.
We then use the Boltzmann equation coherent scattering rates to derive
a set of modified cross sections in section~\ref{sub:MC-scattering},
and identify the approximation made in Wojcik and Tachiya's original
method \cite{Wojcik2002}. In section~\ref{sub:MC-fields}, we describe
a new method that we have developed for performing simulations in
spatially-varying electric fields. The model system that we are studying
is extensively described in section~\ref{sec:model_description},
and finally, we present our results in section~\ref{sec:results},
including comparisons with the results from both our implementation
of Wojcik and Tachiya's method, as well as an independent Boltzmann
equation solution.

\section{Theory}

\subsection{Coherent scattering\label{sub:Coherent-elastic-scattering}}

Designing simulations of swarm transport in liquids and dense gases
presents additional challenges compared to the ideal gas case. Because
the inter-particle spacing of the neutral particles is often less
than the de Broglie wavelength of the swarm particles, the swarm particles
must often interact with several neutral particles at the same time,
which means that any spatial or temporal correlations between said
particles will have an effect on scattering events. The Cohen-Lekner
theory of electron transport \cite{Cohen1967} describes these effects
in terms of two rates of transfer -- momentum and energy -- that occur
independently.

Cohen and Lekner express the electron distribution function in a basis
of spherical harmonics. They then modify the standard Boltzmann collision
integral to include the dynamic structure factor S$\left(\mathbf{\Delta k,}\omega\right)$,
as motivated by van Hove's definition of the ensemble cross section~\cite{VanHove1954},
and then show that when the necessary integrals have been performed,
the dependence is only upon the static structure factor.

Upon solving the equations for the time evolution of the distribution
function, they ascribe a physical meaning to two of the mean free
path lengths that appear in the collision integral expansion. The
first fully determines the rate of energy transferred from the swarm
particles. It is independent of the structure of the medium, and is
given by the mean free path corresponding to single-particle elastic
scattering:

\begin{equation}
\Lambda_{0}=(n_{0}\sigma_{m})^{-1}=\left(n_{0}2\pi\int_{0}^{\pi}d\chi\sin\chi\left(1-\cos\chi\right)\sigma_{sp}\left(\epsilon,\chi\right)\right)^{-1},\label{eq:mfp_energy}
\end{equation}
where $n_{0}$ is the number density of the neutral molecules, $\sigma_{sp}\left(\epsilon,\chi\right)$
is the angle-differential elastic cross section for scattering with
a single particle (also known as the binary or gas-phase cross section),
and $\sigma_{m}$ is the usual definition of the momentum transfer
cross section in the absence of coherent effects. Throughout the present
work, $\epsilon$ refers to the relative energy in the centre-of-mass
frame during a collision, and $\chi$ represents the angle through
which the relative velocity is changed.

The second mean free path partly includes the effect of the medium
and contains all information about the rate at which momentum is transferred:

\begin{equation}
\Lambda_{1}=(n_{0}\tilde{\sigma}_{m})^{-1}=\left(n_{0}2\pi\int_{0}^{\pi}d\chi\sin\chi\left(1-\cos\chi\right)\sigma_{sp}\left(\epsilon,\chi\right)S\left(\Delta\mathbf{k}\right)\right)^{-1},\label{eq:mfp_momentum}
\end{equation}
where $S\left(\Delta\mathbf{k}\right)$ is the static structure factor
as a function of the momentum transferred and $\tilde{\sigma}_{m}$
represents a structure modification of the momentum transfer cross
section.

In a recent paper \cite{Boyle2012}, the explicit rates of energy
and momentum transfer were calculated with the inclusion of structure
in the Boltzmann equation. The components of this transfer due to
the collision term, in the case of zero temperature, are:
\begin{align}
\frac{d}{dt}\langle n\, m\bm{v}\rangle\Bigg|_{\mathrm{coll}} & =-\langle n_{0}v\tilde{\sigma}_{m}(v)m\bm{v}\rangle+O(\omega)+O(\frac{m}{m_{0}})\nonumber \\
 & =-\langle v\Lambda_{1}^{-1}(v)m\bm{v}\rangle\label{eq:Boltzmann_v_balance}
\end{align}
and
\begin{align}
\frac{d}{dt}\langle n\,\epsilon\rangle\Bigg|_{\mathrm{coll}} & =-2\frac{m}{m_{0}}\langle n_{0}v\sigma_{m}(v)\epsilon\rangle+O(\omega^{2})+O\left(\left(\frac{m}{m_{0}}\right)^{2}\right)\nonumber \\
 & =-2\frac{m}{m_{0}}\langle v\Lambda_{0}^{-1}(v)\epsilon\rangle,\label{eq:Boltzmann_eps_balance}
\end{align}
where $\left\langle \,\right\rangle $ represents averaging over velocity
space and $n$ is the number density of the charged particles.

Note that these representative mean free paths should be considered
independently, and should be only thought of as an average rate of
transfer of the relevant quantity, rather than as a prescription for
separate collision events. We define the ratio $\Gamma(\epsilon)\equiv\Lambda_{0}/\Lambda_{1}=\tilde{\sigma}_{m}/\sigma_{m}$.
In the dilute gas case, $\Gamma(\epsilon)=1$, because the static
structure factor of a dilute gas is unity for all momentum transfers.
However, in a structured medium such as a dense gas or a liquid, the
ratio can deviate markedly from unity. If $\Gamma(\epsilon)<1$, there
is noticeably less momentum transfer than in the single-particle scattering
case, which can be interpreted as a preference towards forward scattering
events. In the opposite case of $\Gamma(\epsilon)>1$, more momentum
transfer occurs, which causes the particle to change direction without
losing as much energy as it would in the single-particle scattering. 

In the case of an isotropic single-particle elastic cross section,
$\sigma_{sp}\left(\epsilon,\chi\right)=\frac{1}{4\pi}\sigma_{sp}\left(\epsilon\right)$,
as in the model described in section \ref{sec:model_description},
and the ratio $\Gamma\left(\epsilon\right)$ reduces to:

\begin{equation}
\Gamma\left(\epsilon\right)=\frac{\Lambda_{0}}{\Lambda_{1}}=\frac{1}{2}\int_{0}^{\pi}d\chi\sin\chi\left(1-\cos\chi\right)S\left(\frac{2\left(2m\epsilon\right)^{1/2}}{\hbar}\sin\left(\frac{\chi}{2}\right)\right),\label{eq:angle_integrated_s}
\end{equation}
where we have assumed the static structure factor depends only on
the magnitude of $\Delta\mathbf{k}$, and $\left|\hbar\mathbf{\Delta k}\right|=\hbar\Delta k\approx2\sqrt{2m\epsilon}\sin\frac{\chi}{2}$
in the limit of a small mass ratio $m/m_{0}$. Throughout the present
work, $m$ refers to the mass of each charged particle, and $m_{0}$
the mass of each neutral molecule. This form of $\Gamma\left(\epsilon\right)$
is sometimes called the angle-integrated static structure factor,
$\bar{S}\left(\epsilon\right)$, and it is this form of the structure
factor that is used in several previous works \cite{Wojcik2002,White2011}.

In the case of dilute gases, where $S\left(\Delta\mathbf{k}\right)=1$,
the energy and momentum transfer rates converge, yielding the single-scattering
model in which every energy transfer is accompanied by a momentum
transfer. When the transfer rates differ, however, this theory is
not directly applicable to Monte Carlo modeling because it does not
give a microscopic description of how much energy and momentum is
transferred in each collision between swarm particles and neutral
particles.

\subsection{Sampling coherent scattering in Monte Carlo simulations\label{sub:MC-scattering}}

Our Monte Carlo simulations are built around sampling sets of scattering
cross sections, $\sigma$, that define the probabilities of all interactions
between the charged particles and the medium. Each cross section represents
a single type of scattering process, for example elastic, direct ionization,
or a particular electronic excitation of the neutral. They usually
depend upon the relative speed of the charged particle during a collision,
and on the scattering angle $\chi$. In the case of a cold background
medium, the collision frequency is simply given by:

\begin{equation}
\nu=n_{0}v\sigma_{tot}\left(v\right),\label{eq:nu_definition}
\end{equation}
where $v$ is the speed of the charged particle and $\sigma_{tot}\left(v\right)$
is the integrated sum of all differential cross sections at that speed
\footnote{Note that single-particle cross sections are ideally defined in terms
of the collision energy, which depends on the relative speed of the
particle in the center of mass frame of the collision. However, in
practice, most measured cross sections are, by necessity, the integral
of the cross section of all possible relative speeds, weighted by
their frequency according to the neutral particle's velocity distribution.
The differences are very small for electrons impacting on even the
lightest atoms, although they may be significant for ion scattering
at very low energies.%
}. This quantity is used to stochastically sample the time between
each collision (further described in section \ref{sub:MC-fields}).
When a collision is simulated, a specific cross section is randomly
selected according to the relative probabilities of the available
cross sections \cite{Brennan1990b}. In the case of single-scattering
collisions with independent gas molecules, the amount of energy and
momentum transferred is fully determined by the initial energy and
the scattering angles.

\begin{figure}
\includegraphics[width=0.9\columnwidth]{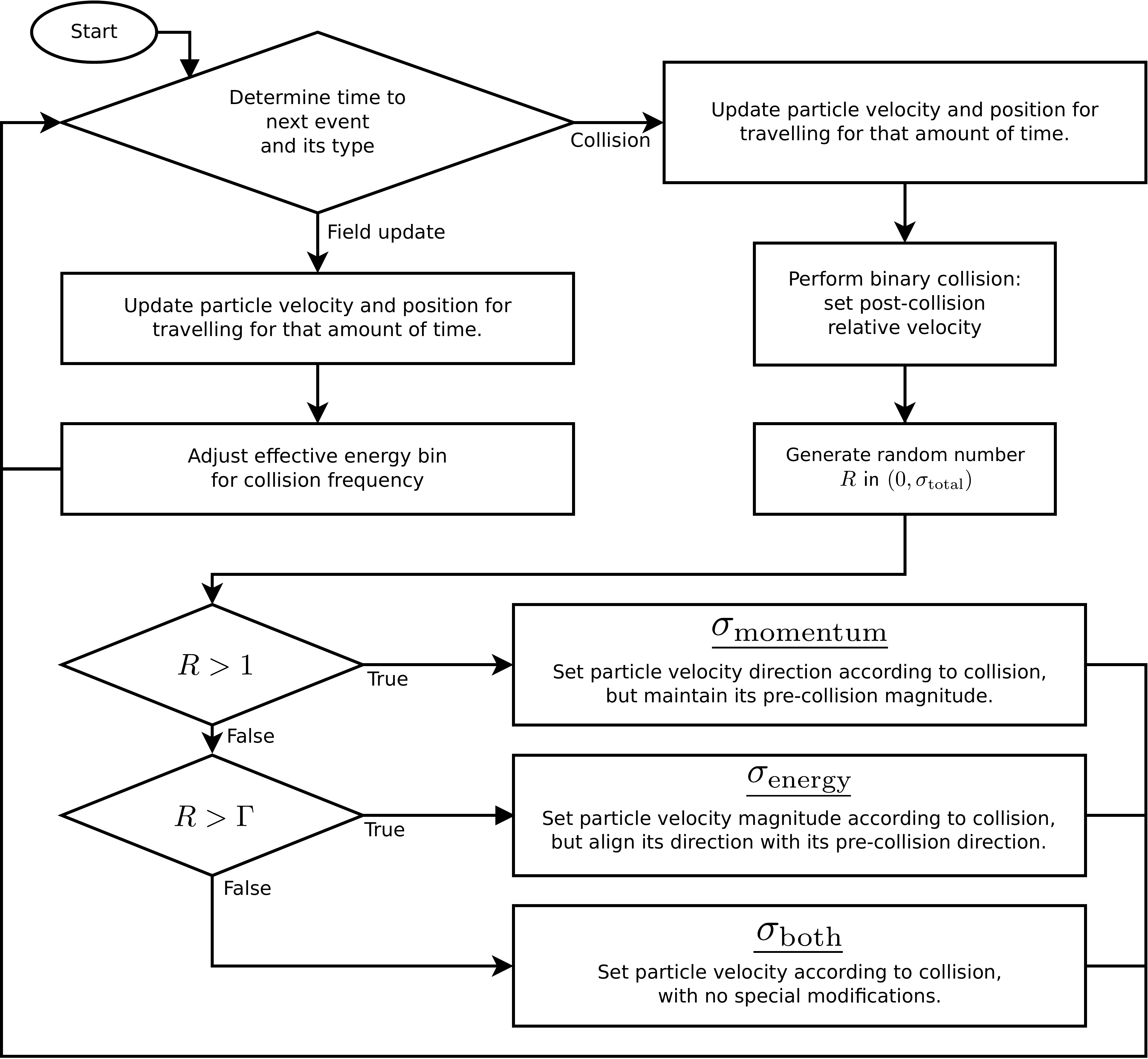}

\protect\caption{\label{fig:flowchart}Flowchart detailing how electric fields and
coherent scattering are implemented in the SSMC code.}
\end{figure}
\begin{figure}
\includegraphics[width=0.9\columnwidth]{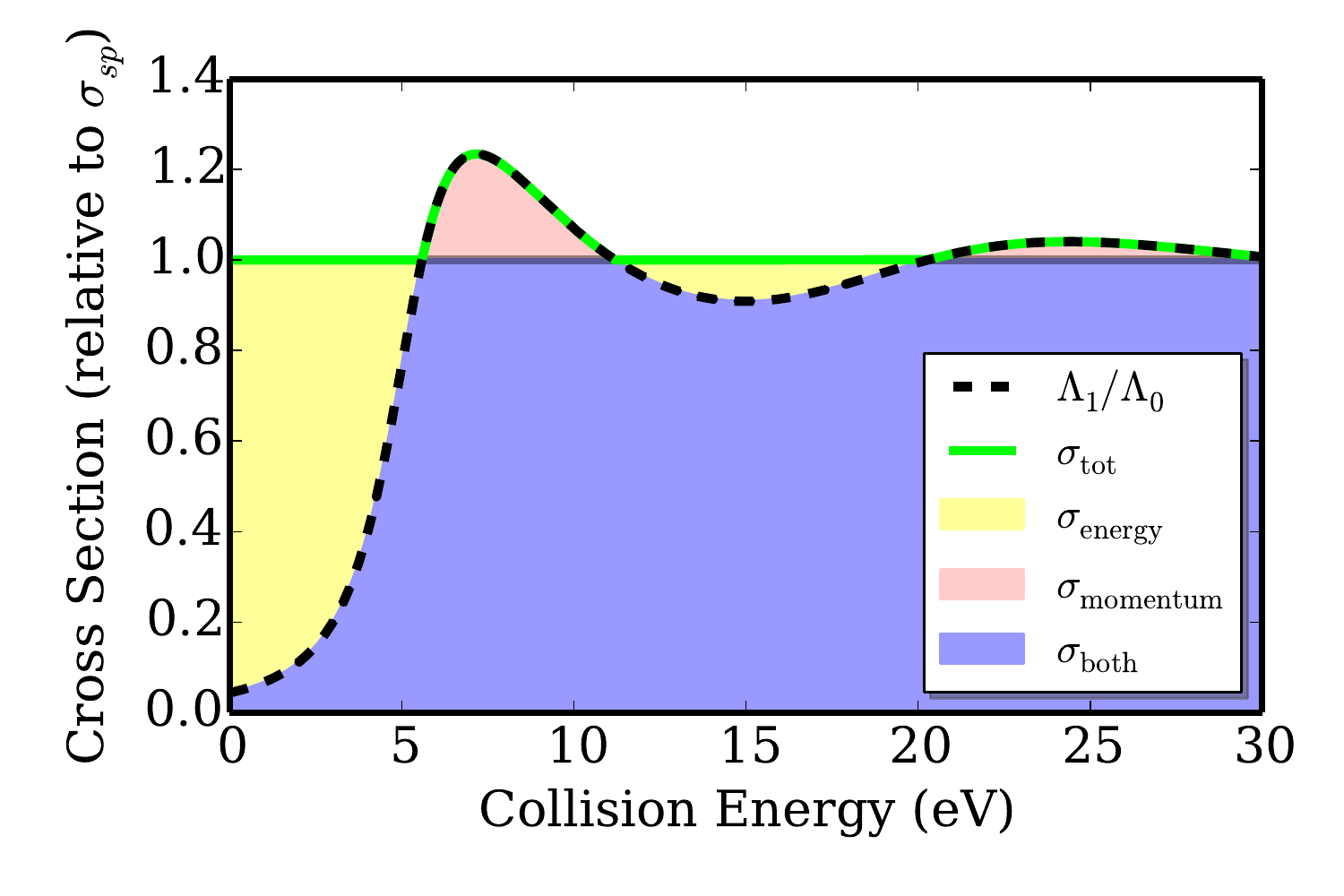}

\protect\caption{\label{fig:schematic}Schematic diagram of the various elastic cross-sections
used in simulating a Percus-Yevick liquid ($\phi=0.4$). All quantities
are given relative to the elastic cross-section for a single particle.
Note that the $\sigma_{\mathrm{tot}}\geq\sigma_{\mathrm{sp}}$.}
\end{figure}

For structured materials, an approximate theory has been developed
by Wojcik and Tachiya \cite{Wojcik2002}, who propose a mechanistic
model of electron transport in rare gas liquids. In what follows,
we have extended this model to be more generally applicable to other
systems, highlighting the approximations and associated errors of
Wojcik and Tachiya's model.

The presence of structure requires the introduction of additional
microscopic processes that, at a macroscopic level, produce the same
rate of energy and of momentum transfer as in the Boltzmann equation
formalism detailed in section \ref{sub:Coherent-elastic-scattering}.
We choose to do this by separating the original, single-particle elastic
cross section into three different processes depending on the ratio
$\Gamma(\epsilon)$, as illustrated in Fig.~\ref{fig:schematic}.
These processes have cross sections, labeled by which quantities are
affected in the collision: $\sigma_{\mathrm{both}}$, $\sigma_{\mathrm{momentum}}$
and $\sigma_{\mathrm{energy}}$. The result of a collision from process
$\sigma_{\mathrm{both}}$ is identical to that of a regular single-particle
scattering collision. For $\sigma_{\mathrm{energy}}$, we start with
a regular single-particle scattering collision, but set the post-collision
direction of motion for the particle to be unchanged. This has the
effect of transferring a minimal amount of momentum whilst maintaining
the same energy transfer as in $\sigma_{\mathrm{both}}$. For $\sigma_{\mathrm{momentum}}$,
we perform a regular single-particle scattering collision, but scale
the post-collision particle speed to be equal to that before the collision.
This results in exactly zero transfer of energy, but some change of
vector momentum.

The path lengths $\Lambda_{0}$ and $\Lambda_{1}$ in section~\ref{sub:Coherent-elastic-scattering}
correspond to transfer rates of $\nu_{m}=vn_{0}\sigma_{m}=v/\Lambda_{0}$
and $\tilde{\nu}_{m}=vn_{0}\tilde{\sigma}_{m}=v/\Lambda_{1}$ for
energy and momentum respectively, where $v$ is the speed of the charged
particle. To achieve these rates, we combine the cross sections in
various ratios depending on the value of $\Gamma(\epsilon)=\Lambda_{0}/\Lambda_{1}$.
If $\Gamma(\epsilon)<1$ we wish to decrease the rate of momentum
transfer, while maintaining energy transfer, and so choose $\sigma_{\mathrm{both}}^{\Gamma<1}=\Gamma(\epsilon)\sigma_{\mathrm{sp}}$,
$\sigma_{\mathrm{energy}}^{\Gamma<1}=(1-\Gamma(\epsilon))\sigma_{\mathrm{sp}}$
and $\sigma_{\mathrm{momentum}}^{\Gamma<1}=0$. In the opposite case,
$\Gamma(\epsilon)>1$, we achieve an increased rate of momentum transfer,
by setting $\sigma_{\mathrm{both}}^{\Gamma>1}=\sigma_{\mathrm{sp}}$,
$\sigma_{\mathrm{momentum}}^{\Gamma>1}=(\Gamma(\epsilon)-1)\sigma_{\mathrm{sp}}$
and $\sigma_{\mathrm{energy}}^{\Gamma>1}=0$. This gives a total elastic
cross section of $\sigma_{\mathrm{tot}}=\max(1,\Gamma(\epsilon))\sigma_{\mathrm{sp}}$.
The complete Monte Carlo procedure, which we refer to as the ``Static
Structure Monte Carlo'' (SSMC) method, is shown as a flowchart in
Fig.~\ref{fig:flowchart}.

The procedure outlined above is designed to reproduce the rates of
energy and momentum transfer in equations (\ref{eq:mfp_energy}) and
(\ref{eq:mfp_momentum}). However, it is not obvious that our construction
of the microscopic processes achieves this goal. In Appendix \ref{sec:correction_appendix},
we show that such a sampling process involving these cross sections
does indeed satisfy these requirements, to within the order of the
mass ratio $m/m_{0}$ as mentioned earlier. These differences are
small enough that they are unlikely to effect electron-atom simulations,
though they may be significant in systems where ions serve as the
charged particles.

Wojcik and Tachiya~\cite{Wojcik2002} studied liquid argon according
to the method described above, but their mechanistic model, hereafter
referred to as the WT method, effectively capped the value of $\Gamma\left(\epsilon\right)$
such that it never exceeded unity. This meant that their total collision
frequency was unaltered from the single-particle scattering case,
and simulation of the particles in the energy regions where $\Gamma\left(\epsilon\right)$
exceeded 1 could only be considered approximately accurate. The difference
between the SSMC and WT methods \cite{Wojcik2002} is shown in Fig.~\ref{fig:schematic},
where the regions labeled $\sigma_{\mathrm{momentum}}$ are absent
in their model, and the total cross section modified accordingly,
so that it is simply $\sigma_{sp}$. For the aforementioned study
of liquid argon, such modifications were only required in a small
energy range for the structure factor that they employed. One of the
purposes of the present study is to determine how this approximation
affects the results for a benchmark Percus-Yevick model, where the
approximation is more significant.

\subsection{Precise treatment of electric fields in Monte Carlo simulations\label{sub:MC-fields}}

Electric fields present a particular challenge for this style of Monte
Carlo simulation. As the collision frequency $\nu$ of a given charged
particle is dependent on its energy $\epsilon$ (see equation~\ref{eq:nu_definition}),
the time between collisions $\tau$ can be altered by the change of
energy of the particle due to the electric field, even as it is undergoing
the transport between collisions. Mathematically, the probability
of a time between collisions greater than $\tau$ can be expressed
as \cite{Skullerud1968}
\begin{equation}
P\left(\tau\right)=\exp\left(-\int_{0}^{\tau}\nu\left(\epsilon\left(t\right)\right)dt\right),\label{eq:time_to_collision}
\end{equation}
where the charged particle's energy is time dependent due to the particle's
passage through an electric field. Explicitly performing this integral
for every collision would be very computationally expensive, given
that the changes in $\epsilon$ will depend on the velocity at which
the particle is traveling and, for non-uniform electric fields, the
position of the charged particle as well. 

One popular approach uses the method of ``null collisions'' \cite{Skullerud1968},
where the collision frequency is calculated based on the maximum collision
frequency $\nu_{0}$ that the particle is likely to be able to reach
during its transport. Using this constant collision frequency, equation
(\ref{eq:time_to_collision}) can be solved by equating $P\left(\tau\right)$
with a uniformly distributed random number $R$, in which case
\begin{equation}
\tau=-\nu_{0}^{-1}\ln R.\label{eq:tau_for_nu_const}
\end{equation}

When the collision occurs, a second random number is generated which
is used to account for this overestimation by allowing the charged
particle to undergo ``null'' collisions, where no exchange of energy
or momentum occurs. This procedure suffers from a requirement to ``backtrack''
if the assumed collision frequency is too low, where it must then
make a second assumption with a higher collision frequency. It is
therefore important to minimize the number of null collisions and
backtracks to optimize the simulation speed, and more modern simulations
\cite{Brennan1991} have been designed with this in mind. The null
collision method effectively amounts to a form of rejection method
for sampling from $P\left(\tau\right)$, which means that potentially
many random numbers are generated for each valid collision.

The simulation presented here uses an alternative approach. The cross-sections
are specified as a function of energy, but are assumed to be constant
within energy bins of width $\delta\epsilon$. These energy bins can
be made arbitrarily small, so there is no loss of accuracy provided
that we are careful to test that the results are independent of the
bin width. However, this means that it is sufficient to recalculate
$\tau$ only when the energy of the particle changes from one bin
to another, so until this occurs, the collision frequency in equation
(\ref{eq:tau_for_nu_const}) remains constant. We therefore design
the simulation so that particles can undergo two types of interactions.
Collisions with neutral particles are described above, and are governed
by $i$, the energy bin of the particle. The second type of interaction
occurs when the energy bin is judged to have changed due to the effect
of the electric field. In such an interaction, the only parameter
that changes is $i$, which either increases or decreases by one,
triggering a recalculation of the time until the next neutral particle
interaction. The time until the change, $t$, is determined by analysis
of the particle's current velocity and the (constant) acceleration
that it is experiencing due to the electric field. This is given by
the smallest real, positive solution to the following equation for
the kinetic energy:

\begin{equation}
\frac{1}{2}m\left(\mathbf{v}_{\mathbf{0}}+\left(\mathbf{a}t\right)\right)^{2}=\epsilon_{i}\pm\frac{1}{2}\delta\epsilon.
\end{equation}
Recalculating the time until collision $\tau$ does not require the
use of another random number. Recalling equation (\ref{eq:tau_for_nu_const}),
we now have additional terms for each change in energy bin:
\begin{eqnarray}
\tau & = & t_{0}+t_{1}+\ldots+t_{n}\nonumber \\
 &  & -\nu_{n}^{-1}\left[\ln\left(R\right)-\nu_{0}\, t_{0}-\nu_{1}\, t_{1}-\ldots-\nu_{n-1}\, t_{n-1}\right],\label{eq:field_time_calc}
\end{eqnarray}
where each $t_{i}$ is determined by the time required for the particle
to change energy from one bin to the next, and each $\nu_{i}$ represents
the corresponding collision frequency for each energy bin at that
time. This equation reduces to equation (\ref{eq:tau_for_nu_const})
in the case of constant $\nu_{n}$ or zero $t{}_{i}$. In practice,
the simulation maintains a running measure of the remaining ``collision
probability'' for each particle. This is the dimensionless quantity
in square brackets in equation \ref{eq:field_time_calc} that is divided
by the current $\nu$ to calculate the time until next collision.

\section{Benchmark System for Modeling Coherent Scattering\label{sec:model_description}}

\subsection{Percus-Yevick Hard Sphere Model\label{sub:PY_model}}

To demonstrate and benchmark the new simulation procedure and code,
we apply it to a simple model system that requires a correct treatment
of structured media. One such model, frequently used in the literature,
is that of a structure for hard-sphere potentials obtained by applying
the Percus-Yevick approximation as a closure to the Ornstein-Zernike
equation, which yields a pair-correlation function \cite{Wertheim1963,Thiele1963},
which in turn can be transformed into a static structure factor via
a Fourier transform and directly used in our simulation. We use the
Verlet and Weiss \cite{Verlet1972} structure factor, which includes
some corrections to better emulate the structure of a real liquid:
\begin{eqnarray}
S\left(\Delta k\right) & = & \Biggl(1+\frac{24\eta}{\Delta k^{2}}\Biggl[\frac{2}{\Delta k^{2}}\frac{12\zeta}{\Delta k^{2}-\beta}\label{eq:py_def}\\
 & + & \frac{\sin\left(\Delta k\right)}{\Delta k}\left(\alpha+2\beta+4\zeta-\frac{24\zeta}{\Delta k^{2}}\right)\nonumber \\
 & + & \frac{2\cos\left(\Delta k\right)}{\Delta k^{2}}\left(\beta+6\zeta-\frac{12\zeta}{\Delta k^{2}}\right)-\alpha-\beta-\zeta\Biggl]\Biggl)^{-1},\nonumber 
\end{eqnarray}
where $\eta=\phi-\frac{\phi^{2}}{16}$, $\alpha=\frac{\left(1+2\eta\right)^{2}}{\left(1-\eta\right)^{4}}$,
$\beta=\frac{-6\eta\left(1+\frac{\eta}{2}\right)^{2}}{\left(1-\eta\right)^{4}}$
and $\zeta=\frac{\eta\alpha}{2}$. This includes a packing density
parameter, $\phi$, which specifies how closely the hard spheres are
packed. It can be written in terms of the hard sphere radius $r$
and the neutral number density $n_{0}$ as $\phi=\frac{4}{3}\pi r^{3}n_{0}$.
This structure factor depends only on the magnitude of the momentum
exchange during a collision. 

\begin{figure}
\includegraphics[width=0.9\columnwidth]{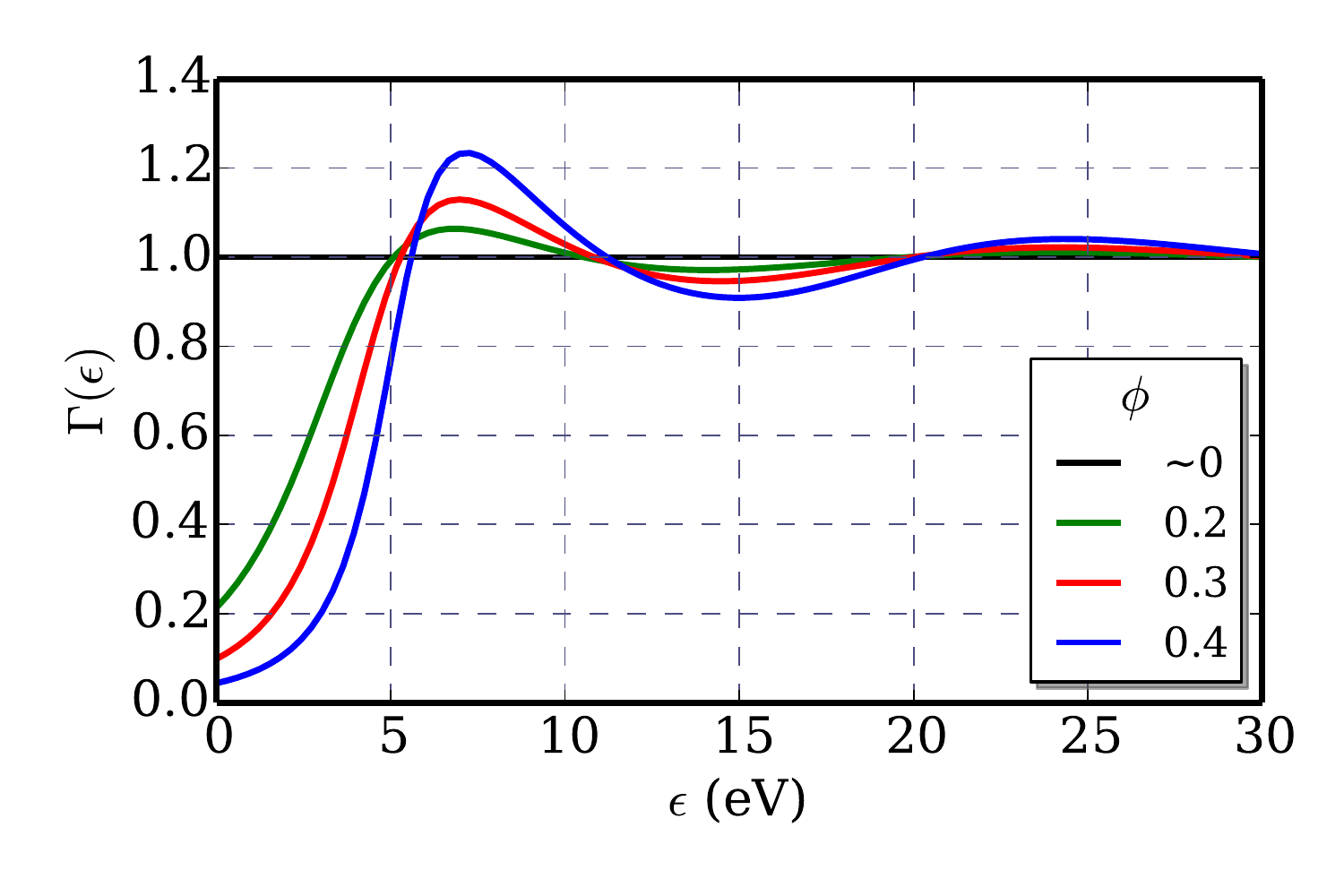}\protect\caption{Angle integrated Percus-Yevick structure factors, from equations (\ref{eq:py_def})
and (\ref{eq:angle_integrated_s}), as a function of particle energy
and volume fraction.\label{fig:py_structure}}
\end{figure}

We have modeled systems with a range of densities, from $\phi\approx0$,
which approximates a dilute gas, to $\phi=0.4$, which states that
$40\%$ of the volume is excluded by the hard-sphere potentials of
the neutral molecules. The angle-integrated forms of each of these
structure factors, as described in equation (\ref{eq:angle_integrated_s}),
are shown in Fig.~\ref{fig:py_structure}.

\subsection{System parameters}

Our Monte Carlo codes use an event-by-event model, where every collision
is considered independently. This allows for a great deal of flexibility
in the specification of scattering mechanics, without any of the approximations
used by ``condensed history'' simulations (see e.g.~\cite{Nahum1999}).
In addition, the swarm approximation --- that all transport particles
are independent --- allows the simulation to be run in parallel. This
makes it ideal for scheduled multi-processor batch jobs, where execution
is not necessarily simultaneous or even sequential.

We have calculated a number of transport coefficients for comparison
with other models. The meaning and derivation of all of these coefficients
are described in \cite{White2011} and \cite{Dujko2008}, but a short
summary is given here. All coefficients are measured as a discrete
function of time-step $t_{i}$. During simulation, if a particle's
history crosses a time-step, its properties (eg: position, velocity,
position squared) at that time are sampled. We choose the z-axis to
be aligned with the electric field, since it is the only element of
the system that has a preferred direction. Each property is added
to a separate running total for each time step, and after the simulation
is complete, the totals are divided by the (in general) time-dependent
total of the number of particles. This results in the average of a
property over all particles, as a function of time. While the transport
coefficients are in general time-varying, for the systems considered
in this study all transport coefficients eventually reach a hydrodynamic
equilibrium after sufficient time has passed. Afterwards, they merely
fluctuate in a small statistical range about the reported equilibrium
value. 

Depending on the required transport coefficients, different properties
of the charged particles must be recorded. In the following table,
each definition \cite{Dujko2008} is of the named property at one
point in time:

\begin{tabular}{|l|l|}
\hline 
Transport coefficient & Definition\tabularnewline
\hline 
\hline 
Mean energy & $\bar{\epsilon}=\left\langle \epsilon\right\rangle $\tabularnewline
\hline 
Bulk drift velocity & $W=\frac{d}{dt}\left\langle r_{z}\right\rangle $\tabularnewline
\hline 
Bulk longitudinal diffusion & $D_{L}=\frac{d}{dt}\left(\left\langle r_{z}^{2}\right\rangle -\left\langle r_{z}\right\rangle ^{2}\right)$\tabularnewline
\hline 
Bulk transverse diffusion & $D_{T}=\frac{1}{2}\sum_{i=x,y}\frac{d}{dt}\left(\left\langle r_{i}^{2}\right\rangle -\left\langle r_{i}\right\rangle ^{2}\right)$\tabularnewline
\hline 
\end{tabular}

\noindent Angle brackets denote an average over all particles, $r_{i}$
represents the appropriate Cartesian coordinate of the position of
the particle and $\epsilon$ represents the energy of the particle. 

In principle, there are two types of transport coefficients, known
as ``flux'' and ``bulk'', which approximately correspond to per-particle
averages and system averages respectively \cite{Dujko2008}. However,
because our model contains no non-conservative collision processes,
the ``flux'' and ``bulk'' quantities should be identical, provided
enough time samples are taken. We have chosen to measure ``bulk''
quantities where possible, because this implictly averages over changes
in velocity between time steps, whereas measuring the flux drift velocity
and diffusion requires sampling instantaneous values at discrete time
values. For the same number of time-steps, without any explicit time-averaging
of velocity, the bulk quantities have far less statistical error.

Two ranges of reduced field strengths were used. An approximately
logarithmic spacing of electric field strengths from $0.001$ to $100$
Td provides a broad picture of the resulting behaviors, while a linear
spacing of field strengths from $2$ to $12$ Td provides detail in
the range in which we expect the two Monte Carlo methods to disagree
most strongly.

We employ the cold-gas limit, in which the neutrals are considered
to be at rest. The present method does not accurately support non-zero
neutral temperatures, as the static structure factor does not contain
information about the temperature of the system. We are presently
formulating a rigorous treatment of non-zero neutral temperatures
using the dynamic structure factor \cite{Tattersall2015}.

All of our simulations were performed at different neutral densities
as prescribed by the volume fraction $\phi$ and the hard-sphere radius
of the single particle cross section $\sigma_{sp}$. The transport
properties presented are independent of the neutral number density,
as it scales inversely with the electric field strength. In all cases,
the hard-sphere cross section for single-particle scattering was set
as $\sigma_{sp}\left(\epsilon\right)=6\,\mathring{A}^{2}$, while
the charged particles were assigned a mass equal to that of an electron,
$m=m_{e}$ and the mass of the neutrals was set to $m_{0}=4\, u$.

\section{Results and discussion\label{sec:results}}

\subsection{Transport coefficients calculated with the new Monte Carlo method.}

\begin{figure}
\includegraphics[width=0.9\columnwidth,height=0.9\textheight,keepaspectratio]{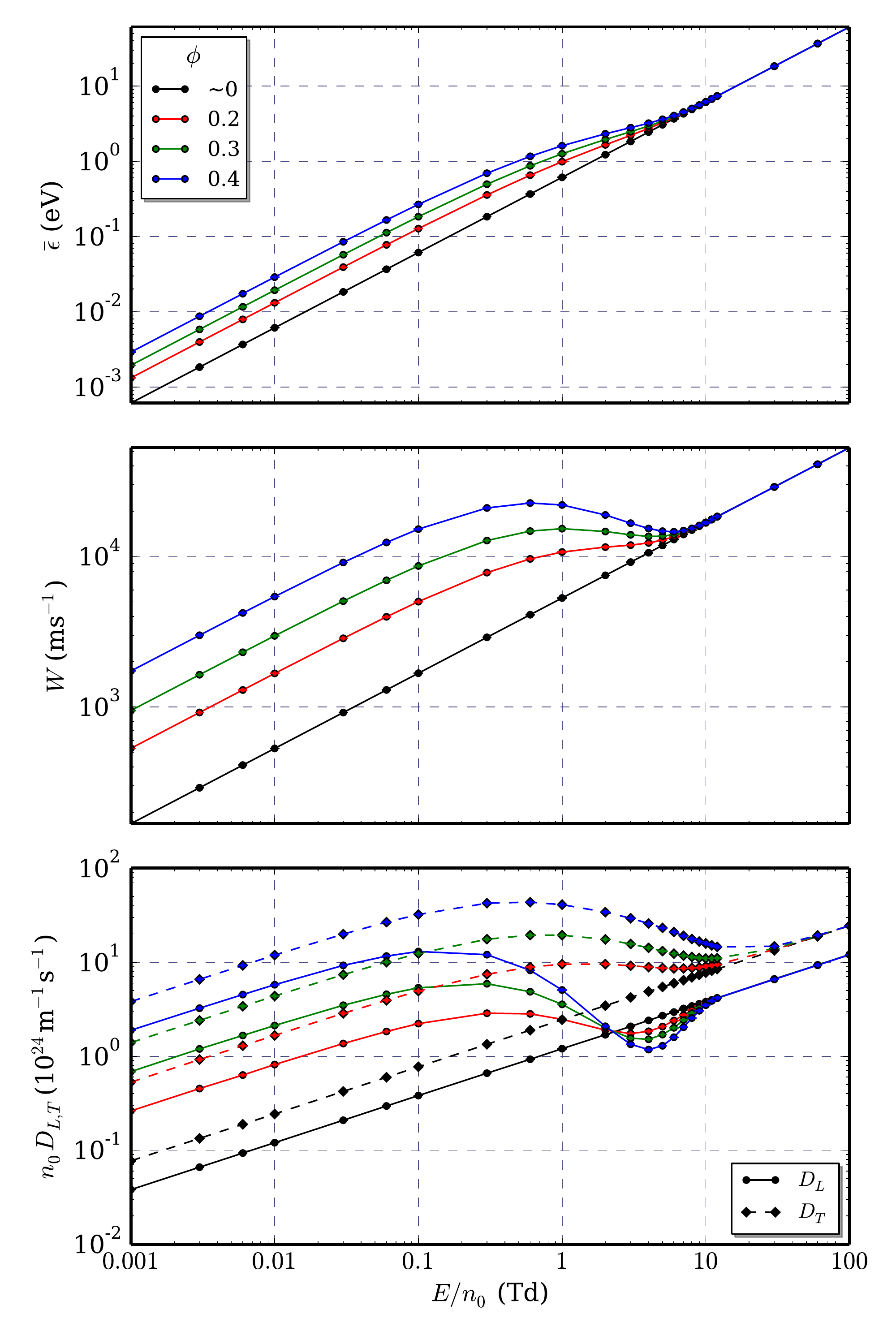}\protect\caption{\label{fig:combined}Mean energy $\epsilon$, drift velocity $W$,
and diffusion coefficients $D_{L}$ and $D_{T}$ for Percus-Yevick
model simulations, as a function of reduced electric field $\mathrm{E/n_{0}}$
and Percus-Yevick packing ratio $\phi$. Error bars are not visible
at this scale.}
\end{figure}

\begin{figure}
\includegraphics[width=0.9\columnwidth]{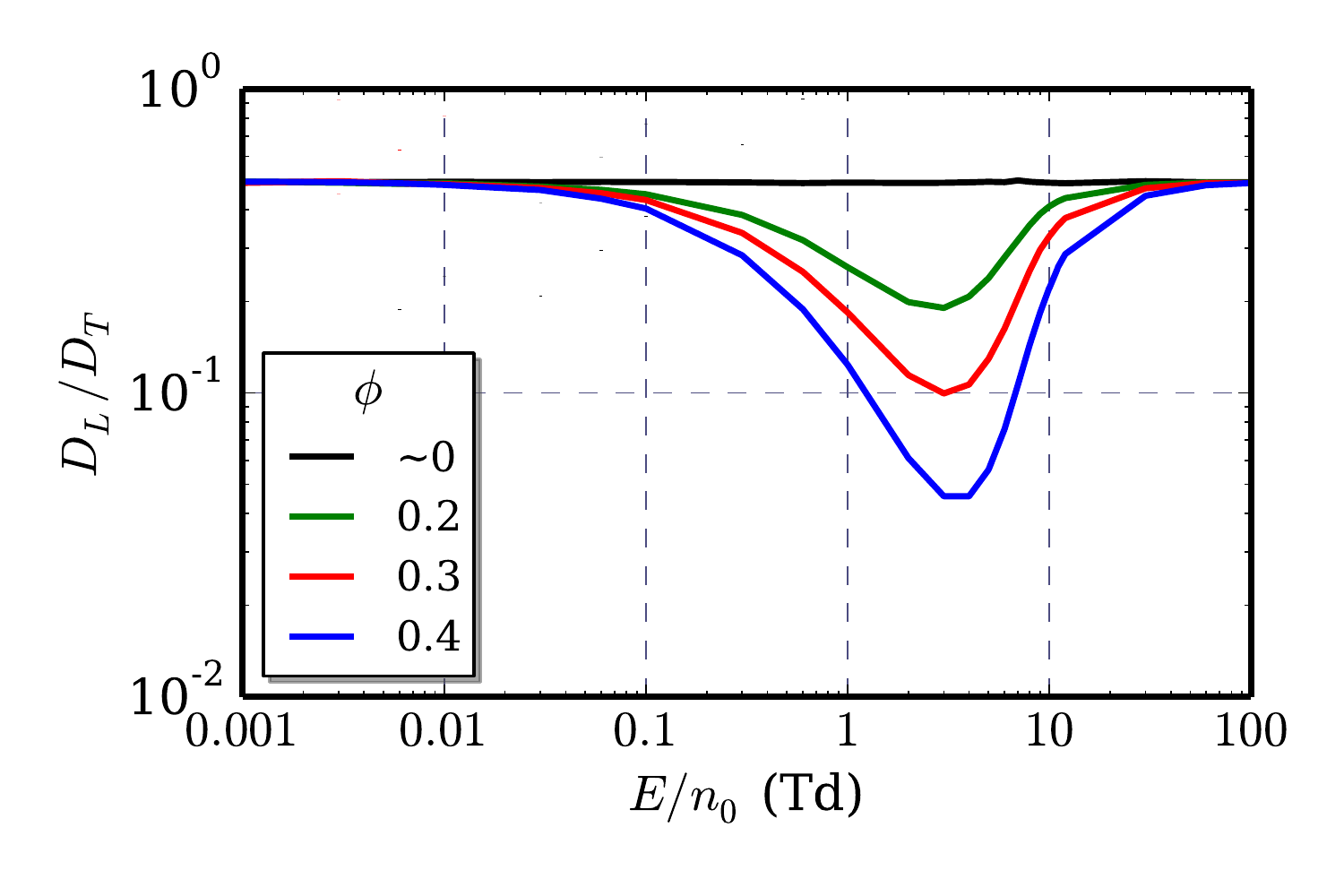}\protect\caption{Ratio of diffusion coefficients as a function of reduced electric
field strength $E/n_{0}$ and packing ratio $\phi$. Note that the
ratio for $\phi\sim0$ is 0.5 as required.\label{fig:diff_ratio}}
\end{figure}

The Percus-Yevick hard-sphere system has been previously studied in
\cite{White2011}, and a comparison with those results provides a
test of our simulation %
\footnote{An error in the calculation of the Percus-Yevick structure factor
used in \cite{White2011} means that the structure factor would be
correct for a fluid where the molecules have a hard-sphere cross-section
of $\sigma=1.5$ Å$^{2}$, not the reported cross-section of $\sigma=6$
Å$^{2}$, although the cross-sections themselves were as reported.
The Boltzmann equation results presented here have been recalculated
with the correct structure factor, as described in section \ref{sub:PY_model}.
The phenomenology reported in \cite{White2011} remains correct, however.%
}. Figure~\ref{fig:combined} shows the various transport coefficients
simulated by the SSMC simulations, using the approach which overcomes
Wojcik and Tachiya's approximation. It is instructive to compare this
figure with Fig.~\ref{fig:py_structure}, to see the field strengths
that are affected most strongly by the features of the structure factors
employed. A table of some of our results is in Appendix \ref{sec:data_table_appendix},
where we have compared them with Boltzmann equation results. The full
dataset is available as supplementary material. We have simulated
enough particles that in general the Monte Carlo statistical error
\cite{Koehler2009a} is not visible at these scales, being less than
$1\%$ in all cases. Agreement with the Boltzmann equation results
is to within 1\% in all cases, so the datasets would not be seen as
distinct if shown in the above figures, but a detailed comparison
is presented in section~\ref{sub:Comparisons-with-Boltzmann}.

The features of the results are discussed in detail in \cite{White2011},
and we will not repeat that discussion in depth here. One key feature
is the presence of structure-induced negative-differential conductivity
(defined in \cite{White2009}) apparent in the drift velocity: at
moderate field strengths of 1-10 Td, the drift velocity is inversely
proportional to the field strength. This is because at low field strengths,
the presence of coherent scattering causes an anisotropy in particle
scattering which allows the particles to be affected more consistently
by the field, raising their velocity in comparison to the structure-free
case. At higher field strengths, the mean particle energy is higher,
leading to a reduced de Broglie wavelength, which means that the charged
particles interact with fewer neutral molecules, so the coherent effects
are reduced. This results in a net reduction of forward motion despite
a higher average energy.

We would also like to highlight the variation in the anisotropic diffusion
as a function of $\phi$. In the case of a hard-sphere gas with no
structure, we expect the ratio $D_{L}/D_{T}=0.5$ \cite{Robson2006},
which, as shown in Fig. \ref{fig:diff_ratio}, is demonstrated by
our simulations. When structure is introduced, this ratio changes
significantly. This effect has been previously explored in \cite{White2011}
and \cite{Boyle2012} through the extended Generalized Einstein Relation.
We note that a multiterm Boltzmann equation solution is required to
achieve the accuracy of the SSMC technique. The SSMC simulations have
no difficulties in accurately representing this anisotropy in the
velocity distribution function.

\subsection{Comparisons with Boltzmann equation and Wojcik and Tachiya's method.\label{sub:Comparisons-with-Boltzmann}}

\begin{figure}
\includegraphics[width=0.9\columnwidth]{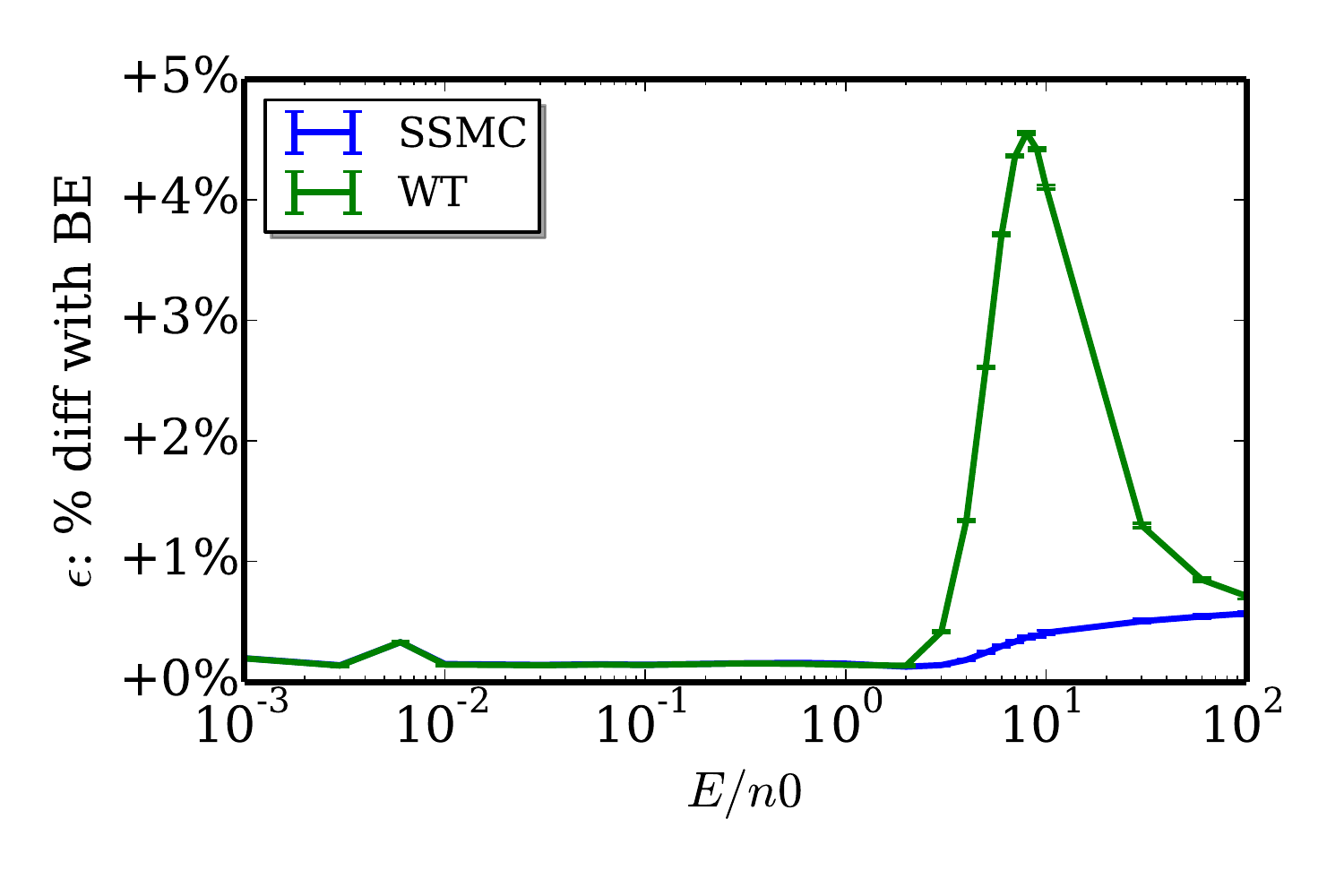}

\includegraphics{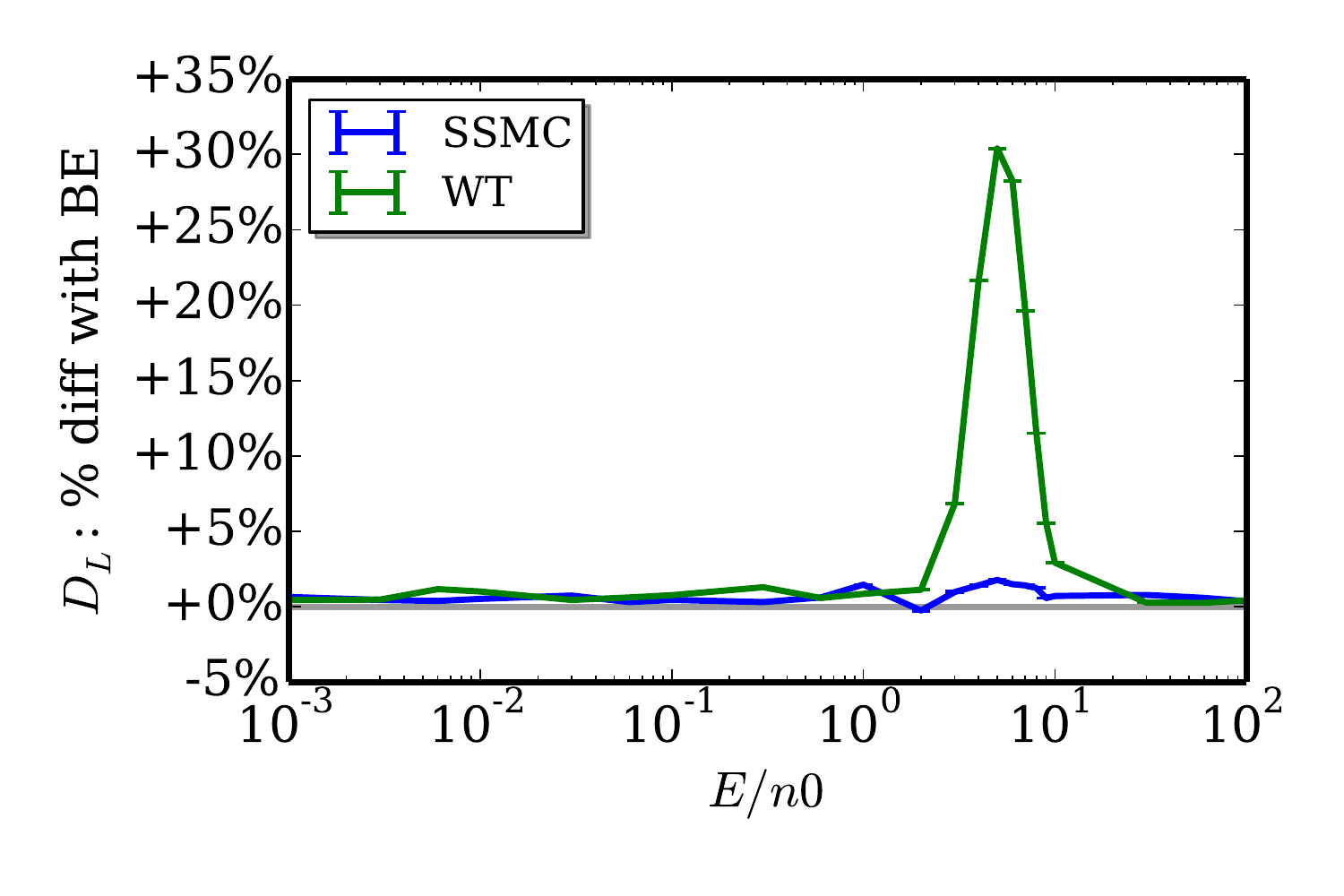}\protect\caption{Mean energy $\epsilon$ and longitudinal diffusion $D_{L}$ percentage
difference for each Monte Carlo model versus the Boltzmann equation
(BE) model, for the Percus-Yevick structure factor at $\phi=0.4$.\label{fig:comparison}}
\end{figure}

In Fig. \ref{fig:comparison}, we present the percentage difference
between our SSMC and WT results and the Boltzmann equation solution.
Our implementation of the WT method shows good agreement over most
regions of field strength, however for some larger field strengths,
errors of up to 5\% in the mean energy and up to 35\% in the diffusion
coefficients become apparent. These differences occur when the energies
of the electrons are within the regions that are truncated by the
WT method. We can identify two competing factors that have an effect
when the particle's energy is in these regions, causing differences
between the SSMC and WT methods. Firstly, the collision frequency
is enhanced, and since every collision has a chance of both removing
some energy from the particle and changing the direction of the particle
away from the direction of the electric field, this means that particles
will tend to lose energy at a greater rate. However, this is balanced
by the presence of the new momentum-only collision, which can occur
up to 28\% of the time in these regions. The presence of such collisions
will tend to decrease the energy transfer rate. Nevertheless, Fig.
\ref{fig:comparison} clearly shows that the combination of the effects
is observable for the Percus-Yevick $\phi=0.4$ case, with a peak
difference at approximately $8$ Td. This corresponds to a mean energy
of about 5 eV (see Fig. \ref{fig:combined}), which is at the peak
of the $\Gamma\left(\epsilon\right)$ function. That peak is where
we would expect the WT approximation to be least suitable. 

For the present model, the disagreements with the Boltzmann equation
results are less than $1\%$ over all field strengths considered.
Such differences are of the order of the numerical schemes used in
the SSMC and Boltzmann equation methods. We suspect the remaining
differences are a result of energy meshes used in the Monte Carlo
codes or Boltzmann equation numerical solutions.

\section{Concluding Remarks}

We have presented a new Monte Carlo simulation code which accurately
accounts for the effects of structure in non-gaseous systems by employing
a modified mechanistic per-collision interpretation of the Cohen and
Lekner method for solving the Boltzmann equation. The SSMC results
accurately replicate those calculated via a multi-term solution of
the Boltzmann equation to within 1\%, with agreement significantly
improved over those results obtained using the WT method, where errors
of up to 35\% were observed in some transport coefficients. Future
work will be focused on utilizing a dynamic structure factor to account
for other structural and collective effects, including non-zero background
temperatures.

\clearpage{}

\bibliographystyle{apsrev4-1}
\bibliography{library}

\appendix

\section{Comparison between Monte-Carlo and Boltzmann equation averages\label{sec:correction_appendix}}

As discussed in section~\ref{sub:MC-scattering}, the three different
processes in the SSMC have been chosen in order to reproduce the mean
rate of transfer of energy and momentum that was obtained in the derivation
of the Boltzmann equation. These different processes, corresponding
to the ``cross sections'' defined in the main text, are: $\sigma_{\mathrm{both}}$,
a normal collision that exchanges both energy and momentum; $\sigma_{\mathrm{energy}}$,
a collision without direction change that aims to allow only energy
exchange; and $\sigma_{\mathrm{momentum}}$, a collision with only
momentum exchange. In this appendix we show that, for isotropic scattering
from stationary neutrals (i.e. we consider only $T=0$ as in accordance
with the regime of validity of the Monte-Carlo simulations), these
processes give rise to the same rates of energy and momentum transfer
as the Boltzmann equation, after neglecting terms dependent on the
mass-ratio $m/m_{0}$. Note that in this appendix, dashes refer to
post-collision quantities.

In a collision, it is appropriate to consider the particle in the
center of mass frame. In this frame, isotropic scattering implies
that the particle's relative velocity $\bm{v}_{\mathrm{rel}}=\bm{v}-\bm{g}$,
where $\bm{g}$ is the velocity in the center of mass frame, is unchanged
in magnitude ($|v_{\mathrm{rel}}^{\prime}|=|v_{\mathrm{rel}}|$) and
with a randomly assigned angle such that all values of $\cos\chi$,
where $\chi$ is the scattering angle, are equally likely. We assume,
without loss of generality, that the direction of the particle's velocity
prior to the collision is aligned with the $z$ axis. Note that we
do not need to consider the electric field in this discussion.

For a normal (``both'') collision, one can easily show \cite{Robson2006}
that scattering from a stationary neutral leads to an angle-dependent
energy change given by
\begin{equation}
\Delta\epsilon_{\mathrm{both}}=\epsilon\frac{2mm_{0}}{(m+m_{0})^{2}}(\cos\chi-1),
\end{equation}
and a change of the $z$-component of the momentum given by
\begin{equation}
\Delta v_{z,\mathrm{both}}=\frac{m_{0}}{m+m_{0}}v(\cos\chi-1).
\end{equation}
Hence the average transfer of energy is 
\begin{eqnarray}
\langle\Delta\epsilon\rangle_{\mathrm{both}} & = & \frac{1}{2}\int_{-1}^{1}\Delta\epsilon_{\mathrm{both}}\, d\cos\chi\\
 & = & 2mm_{0}\epsilon/(m+m_{0})^{2}
\end{eqnarray}
and the average transfer of momentum, for which the components perpendicular
to $z$ average to zero, is 
\begin{equation}
\langle\Delta v_{z}\rangle_{\mathrm{both}}=-m_{0}v_{z}/(m+m_{0}).\label{eq:deltav_both}
\end{equation}

For an ``energy-only'' collision, we select a random change in energy
through a random ``false'' angle $\chi$, and calculate the energy
change as if a normal collision through this angle had occurred. This
preserves the average transfer of energy $\langle\Delta\epsilon\rangle_{\mathrm{energy}}=\langle\Delta\epsilon\rangle_{\mathrm{both}}$.
We then discard the angle $\chi$, resetting the direction of motion
of the particle to that before the collision. This has the unfortunate
side effect of producing a change in momentum, contrary to the purpose
of producing a collision with no momentum transfer. The change of
momentum in this case is simply given by:
\begin{align}
\Delta v_{z,\mathrm{energy}} & =\sqrt{\frac{2}{m}}\left(\sqrt{\epsilon-\Delta\epsilon(\chi)}-\sqrt{\epsilon}\right)\nonumber \\
 & =v\left(\sqrt{1+\frac{2mm_{0}}{(m+m_{0})^{2}}(\cos\chi-1)}-1\right)
\end{align}
and upon averaging over $\cos\chi$ we find:
\begin{align*}
\langle\Delta v_{z}\rangle_{\mathrm{energy}} & =-v\left[1-\frac{1}{3a}\left(1-(1-2a)^{3/2}\right)\right]\\
 & =-\frac{m\left(3m_{0}-m\right)v_{z}}{3m_{0}(m+m_{0})}
\end{align*}
where $a=2mm_{0}/(m+m_{0})^{2}$, and the second result is arrived
at after some algebraic manipulations, under the assumption that $m<m_{0}$.

For a ``momentum only'' collision, we perform a collision as normal
using an angle $\chi$ but set the post-collision energy to the pre-collision
energy. Hence, it is obvious that $\mbox{\ensuremath{\langle\Delta\epsilon\rangle_{\mathrm{momentum}}}=0}$
but the modification of the energy also has an impact on the average
momentum change, which we calculate below. Note that $\bm{v}^{\prime}=v\hat{\bm{v}}^{\prime}$,
where $\hat{\bm{v}}^{\prime}$ is identical to that of a normal collision.
Hence,
\begin{equation}
\bm{v}_{\mathrm{both}}^{\prime}\cdot\hat{\bm{z}}=\frac{v}{m+m_{0}}\left(m_{0}\cos\chi+m\right)
\end{equation}
implies that 
\begin{align}
\hat{\bm{v}}^{\prime}\cdot\hat{\bm{z}} & =\frac{\bm{v}_{\mathrm{both}}^{\prime}\cdot\hat{\bm{z}}}{|\bm{v}_{\mathrm{both}}^{\prime}|}\nonumber \\
 & =\frac{\cos\chi+\frac{m}{m_{0}}}{\sqrt{1+2\frac{m}{m_{0}}\cos\chi+\frac{m^{2}}{m_{0}^{2}}}}
\end{align}
From this, we find the change in the $z$-component of momentum of
a single collision
\begin{align}
\Delta v_{z,\mathrm{momentum}} & =(\bm{v}^{\prime}-\bm{v})\cdot\hat{\bm{z}}=v(\hat{\bm{v}}^{\prime}\cdot\hat{\bm{z}}-1)
\end{align}
and hence obtain an average momentum transfer of 
\begin{align}
\langle\Delta v_{z}\rangle_{\mathrm{momentum}} & =\frac{v}{2}\int_{-1}^{1}\left(\frac{\cos\chi+\frac{m}{m_{0}}}{\sqrt{1+2\frac{m}{m_{0}}\cos\chi+\frac{m^{2}}{m_{0}^{2}}}}-1\right)\, d\cos\chi\nonumber \\
 & =-v_{z}\left(1-\frac{2}{3}\frac{m}{m_{0}}\right)\label{eq:delta_v_momentum}
\end{align}

We must now combine these averages. In the Monte-Carlo simulation
there are two distinct regimes depending on the value of $\Gamma(\epsilon)$.
If $\Gamma(\epsilon)<1$, then the $\sigma_{\mathrm{both}}$ and $\sigma_{\mathrm{energy}}$
cross sections occur with frequencies such that:
\begin{align}
\makebox[2em][l]{\ensuremath{{\displaystyle \left.\frac{d\langle\epsilon\rangle}{dt}\right|_{\Gamma(\epsilon)<1}}}}\nonumber \\
 & =\nu_{\mathrm{sp}}\Gamma(\epsilon)\langle\Delta\epsilon\rangle_{\mathrm{both}}+\nu_{\mathrm{sp}}(1-\Gamma(\epsilon))\langle\Delta\epsilon\rangle_{\mathrm{energy}}\nonumber \\
 & =\nu_{\mathrm{sp}}\langle\Delta\epsilon\rangle_{\mathrm{both}}\nonumber \\
 & =\nu_{\mathrm{sp}}\frac{2mm_{0}}{(m+m_{0})^{2}}\epsilon\label{eq:Gamma_lt_1_eps}
\end{align}
where $\nu_{\mathrm{sp}}(v)=n_{0}v\sigma_{\mathrm{sp}}(v)$ would
be the collision frequency in the absence of structure effects, and
\begin{align}
\makebox[2em][l]{\ensuremath{{\displaystyle m\left.\frac{d\langle\bm{v}\rangle}{dt}\right|_{\Gamma(\epsilon)<1}}}}\nonumber \\
 & =m\nu_{\mathrm{sp}}\Gamma(\epsilon)\langle\Delta\bm{v}\rangle_{\mathrm{both}}+m\nu_{\mathrm{sp}}(1-\Gamma(\epsilon))\langle\Delta\bm{v}\rangle_{\mathrm{energy}}\nonumber \\
 & \approx-\frac{mm_{0}\nu_{\mathrm{sp}}\bm{v}}{m+m_{0}}\Gamma(\epsilon)+O(\frac{m}{m_{0}}),\label{eq:Gamma_lt_1_v}
\end{align}
where we have generalized the momentum transfer for any initial velocity
direction, i.e. by replacing $v_{z}$ in equations (\ref{eq:deltav_both})
and (\ref{eq:delta_v_momentum}) by $\bm{v}$. In order to compare
to the Boltzmann equations, we first note that $\sigma_{m}(v)=\sigma_{\mathrm{sp}}(v)$
for isotropic scattering and hence $\nu_{\mathrm{sp}}=v\Lambda_{0}^{-1}$
and $\nu_{\mathrm{sp}}\Gamma(\epsilon)=v\Lambda_{1}^{-1}$. Hence,
we can see that these rates of transfer are in agreement with the
results for the Boltzmann equation (\ref{eq:Boltzmann_v_balance})
and (\ref{eq:Boltzmann_eps_balance}), excepting some factors of order
$m/m_{0}$ that are neglected in our simulations and in the derivation
of the Boltzmann equation. If $\Gamma(\epsilon)>1$ then the ``both''
and ``momentum only'' cross sections occur with frequencies such
that:
\begin{align}
\makebox[2em][l]{\ensuremath{{\displaystyle \left.\frac{d\langle\epsilon\rangle}{dt}\right|_{\Gamma(\epsilon)>1}}}}\nonumber \\
 & =\nu_{\mathrm{sp}}\langle\Delta\epsilon\rangle_{\mathrm{both}}+\nu_{\mathrm{sp}}(\Gamma(\epsilon)-1)\langle\Delta\epsilon\rangle_{\mathrm{momentum}}\nonumber \\
 & =\nu_{\mathrm{sp}}\frac{2mm_{0}}{(m+m_{0})^{2}}\epsilon
\end{align}
and
\begin{align}
\makebox[2em][l]{\ensuremath{{\displaystyle m\left.\frac{d\langle\bm{v}\rangle}{dt}\right|_{\Gamma(\epsilon)>1}}}}\nonumber \\
 & =m\nu_{\mathrm{sp}}\langle\Delta v\rangle_{\mathrm{both}}+m\nu_{\mathrm{sp}}(\Gamma(\epsilon)-1)\langle\Delta v\rangle_{\mathrm{momentum}}\nonumber \\
 & \approx-\frac{mm_{0}\nu_{\mathrm{sp}}\bm{v}}{m+m_{0}}\Gamma(\epsilon)+O(\frac{m}{m_{0}})
\end{align}
and again we recover the Boltzmann equation results with negligible
factors of $m/m_{0}$.

We note that it is possible to correct for these differences of the
order of $m/m_{0}$ by modifying the cross sections $\sigma_{\mathrm{energy}}$
and $\sigma_{\mathrm{momentum}}$. However, this would only be necessary
if we considered systems in which the mass ratios were close to unity.

Finally, we comment on the difference between the SSMC and WT methods.
Wojcik and Tachiya \cite{Wojcik2002} had mentioned that, in their
particular calculation, they assumed the structure factor was mostly
smaller than unity, which also implies that $\Gamma(\epsilon)<1$.
They did not deal with values greater than unity, instead choosing
to set $\Gamma(\epsilon)=\min(\Gamma(\epsilon),1)$. This means that
the rates of energy and momentum transfer are unchanged when $\Gamma(\epsilon)<1$
and continue to follow equations (\ref{eq:Gamma_lt_1_eps}) and (\ref{eq:Gamma_lt_1_v}).
However, for $\Gamma(\epsilon)>1$, this leads to an error in the
WT method in the momentum transfer which is of the order:
\begin{equation}
-\frac{mm_{0}\nu_{\mathrm{sp}}v}{m+m_{0}}(\Gamma(\epsilon)-1).
\end{equation}
There is no error in the energy transfer using the WT method.

\section{Benchmark Values for Transport Coefficients\label{sec:data_table_appendix}}

\begin{sidewaystable*}
\begin{center}\begin{tabular}{|c|l|l|l|l|l|l|l|} \hline
\multirow{2}{*}{$\phi$} & \multicolumn{1}{c|}{$E/n_{0}$} & \multicolumn{3}{c|}{Mean energy (eV)} & \multicolumn{3}{c|}{Drift Velocity ($\textrm{ms}^{-1}$)} \tabularnewline
\cline{3-8}
 & \multicolumn{1}{c|}{(Td)} &
\multicolumn{1}{c|}{BE} & \multicolumn{1}{c|}{SSMC} & \multicolumn{1}{c|}{Diff.} &
\multicolumn{1}{c|}{BE} & \multicolumn{1}{c|}{SSMC} & \multicolumn{1}{c|}{Diff.} 
\tabularnewline\hline

0.0 & 0.03 & 0.018232 & 0.018338 $\pm$ 0.023\% & 0.58\% & 910.63 & 918.52 $\pm$ 0.0058\% & 0.86\%\tabularnewline
 & 0.3 & 0.18232 & 0.1834 $\pm$ 0.022\% & 0.59\% & 2879.7 & 2905.6 $\pm$ 0.0064\% & 0.90\%\tabularnewline
 & 3 & 1.8232 & 1.8333 $\pm$ 0.02\% & 0.55\% & 9106.3 & 9186.5 $\pm$ 0.0074\% & 0.88\%\tabularnewline
 & 10 & 6.077 & 6.113 $\pm$ 0.022\% & 0.58\% & 16626 & 16783 $\pm$ 0.0096\% & 0.94\%\tabularnewline
 & 30 & 18.232 & 18.34 $\pm$ 0.024\% & 0.58\% & 28797 & 29077 $\pm$ 0.01\% & 0.97\%\tabularnewline
\hline
0.2 & 0.03 & 0.039039 & 0.039149 $\pm$ 0.012\% & 0.28\% & 2851.35 & 2861.91 $\pm$ 0.002\% & 0.37\%\tabularnewline
 & 0.3 & 0.35569 & 0.35667 $\pm$ 0.015\% & 0.27\% & 7797 & 7827.7 $\pm$ 0.0027\% & 0.39\%\tabularnewline
 & 3 & 2.2039 & 2.2102 $\pm$ 0.014\% & 0.28\% & 11860.7 & 11921.1 $\pm$ 0.0066\% & 0.50\%\tabularnewline
 & 10 & 6.112 & 6.141 $\pm$ 0.017\% & 0.48\% & 16726 & 16851.5 $\pm$ 0.0055\% & 0.75\%\tabularnewline
 & 30 & 18.25 & 18.351 $\pm$ 0.023\% & 0.55\% & 28833 & 29071 $\pm$ 0.0048\% & 0.82\%\tabularnewline
\hline
0.3 & 0.03 & 0.057118 & 0.057228 $\pm$ 0.01\% & 0.19\% & 5044.64 & 5055.53 $\pm$ 0.0011\% & 0.21\%\tabularnewline
 & 0.3 & 0.49489 & 0.49585 $\pm$ 0.01\% & 0.19\% & 12756.1 & 12788.1 $\pm$ 0.0016\% & 0.25\%\tabularnewline
 & 3 & 2.4654 & 2.4703 $\pm$ 0.012\% & 0.20\% & 13909 & 13959.3 $\pm$ 0.0047\% & 0.36\%\tabularnewline
 & 10 & 6.127 & 6.152 $\pm$ 0.017\% & 0.41\% & 16756.5 & 16863.1 $\pm$ 0.004\% & 0.63\%\tabularnewline
 & 30 & 18.259 & 18.354 $\pm$ 0.015\% & 0.51\% & 28852.2 & 29065.3 $\pm$ 0.0031\% & 0.73\%\tabularnewline
\hline
0.4 & 0.03 & 0.084858 & 0.084979 $\pm$ 0.0098\% & 0.14\% & 9132.66 & 9139.8 $\pm$ 0.00079\% & 0.08\%\tabularnewline
 & 0.3 & 0.69229 & 0.69337 $\pm$ 0.0083\% & 0.15\% & 21031.8 & 21066.9 $\pm$ 0.0012\% & 0.16\%\tabularnewline
 & 3 & 2.7924 & 2.7963 $\pm$ 0.0065\% & 0.14\% & 16637.7 & 16680.3 $\pm$ 0.0038\% & 0.25\%\tabularnewline
 & 10 & 6.14 & 6.165 $\pm$ 0.018\% & 0.40\% & 16767.6 & 16871.6 $\pm$ 0.0033\% & 0.62\%\tabularnewline
 & 30 & 18.269 & 18.361 $\pm$ 0.017\% & 0.50\% & 28872.5 & 29076.6 $\pm$ 0.0028\% & 0.70\%\tabularnewline

\hline

\end{tabular}\end{center}

\protect\caption{Benchmark values for transport coefficients: mean energy and drift
velocity. Boltzmann equation (BE) results are compared with the SSMC
results. Quoted error is twice the standard error in the mean of the
SSMC results. Comparisons between BE and SSMC are shown as a third
sub-column for each type of transport coefficient.}
\end{sidewaystable*}

\begin{sidewaystable*}
\begin{center}\begin{tabular}{|c|l|l|l|l|l|l|l|} \hline
\multirow{2}{*}{$\phi$} & \multicolumn{1}{c|}{$E/n_{0}$} &
\multicolumn{3}{c|}{$n_0 D_L$ (10${}^{24}$ m${}^{-1}$s${}^{-1}$)} & 
\multicolumn{3}{c|}{$n_0 D_T$ (10${}^{24}$ m${}^{-1}$s${}^{-1}$)} \tabularnewline
\cline{3-8}
& \multicolumn{1}{c|}{(Td)} &
\multicolumn{1}{c|}{BE} & \multicolumn{1}{c|}{SSMC} & \multicolumn{1}{c|}{Diff.} & 
\multicolumn{1}{c|}{BE} & \multicolumn{1}{c|}{SSMC} & \multicolumn{1}{c|}{Diff.}  
\tabularnewline\hline

0.0 & 0.03 & 0.20726 & 0.20859 $\pm$ 0.01\% & 0.64\% & 0.42195 & 0.4224 $\pm$ 0.0041\% & 0.10\%\tabularnewline
 & 0.3 & 0.65543 & 0.66011 $\pm$ 0.011\% & 0.71\% & 1.33432 & 1.33984 $\pm$ 0.006\% & 0.41\%\tabularnewline
 & 3 & 2.0726 & 2.0767 $\pm$ 0.014\% & 0.19\% & 4.2195 & 4.2261 $\pm$ 0.0059\% & 0.15\%\tabularnewline
 & 10 & 3.7841 & 3.8073 $\pm$ 0.012\% & 0.61\% & 7.7037 & 7.7459 $\pm$ 0.0067\% & 0.54\%\tabularnewline
 & 30 & 6.5543 & 6.6398 $\pm$ 0.012\% & 1.30\% & 13.3432 & 13.3487 $\pm$ 0.0059\% & 0.04\%\tabularnewline
\hline
0.2 & 0.03 & 1.35966 & 1.36532 $\pm$ 0.006\% & 0.41\% & 2.8418 & 2.8603 $\pm$ 0.0091\% & 0.65\%\tabularnewline
 & 0.3 & 2.8614 & 2.8671 $\pm$ 0.0092\% & 0.19\% & 7.4604 & 7.4461 $\pm$ 0.0032\% & -0.19\%\tabularnewline
 & 3 & 1.7277 & 1.7446 $\pm$ 0.022\% & 0.97\% & 9.1902 & 9.1665 $\pm$ 0.0066\% & -0.25\%\tabularnewline
 & 10 & 3.677 & 3.7075 $\pm$ 0.018\% & 0.82\% & 9.0054 & 9.0232 $\pm$ 0.0048\% & 0.19\%\tabularnewline
 & 30 & 6.5411 & 6.5948 $\pm$ 0.012\% & 0.82\% & 13.6167 & 13.6324 $\pm$ 0.0042\% & 0.11\%\tabularnewline
\hline
0.3 & 0.03 & 3.4743 & 3.4834 $\pm$ 0.0071\% & 0.26\% & 7.3668 & 7.3715 $\pm$ 0.0086\% & 0.06\%\tabularnewline
 & 0.3 & 5.8751 & 5.9043 $\pm$ 0.0076\% & 0.49\% & 17.561 & 17.571 $\pm$ 0.0048\% & 0.06\%\tabularnewline
 & 3 & 1.5321 & 1.5553 $\pm$ 0.018\% & 1.50\% & 15.651 & 15.626 $\pm$ 0.0074\% & -0.15\%\tabularnewline
 & 10 & 3.5985 & 3.5869 $\pm$ 0.023\% & -0.32\% & 11.029 & 10.952 $\pm$ 0.0096\% & -0.70\%\tabularnewline
 & 30 & 6.5336 & 6.5951 $\pm$ 0.0065\% & 0.94\% & 13.985 & 13.991 $\pm$ 0.0072\% & 0.04\%\tabularnewline
\hline
0.4 & 0.03 & 9.182 & 9.2498 $\pm$ 0.0041\% & 0.73\% & 19.8166 & 19.8841 $\pm$ 0.004\% & 0.34\%\tabularnewline
 & 0.3 & 11.9922 & 12.0287 $\pm$ 0.0071\% & 0.30\% & 42.211 & 42.43 $\pm$ 0.0069\% & 0.51\%\tabularnewline
 & 3 & 1.3271 & 1.34 $\pm$ 0.027\% & 0.97\% & 29.2923 & 29.3425 $\pm$ 0.0017\% & 0.17\%\tabularnewline
 & 10 & 3.4836 & 3.5084 $\pm$ 0.0072\% & 0.71\% & 15.81 & 15.813 $\pm$ 0.0067\% & 0.02\%\tabularnewline
 & 30 & 6.526 & 6.5769 $\pm$ 0.0062\% & 0.77\% & 14.7608 & 14.7684 $\pm$ 0.0038\% & 0.05\%\tabularnewline

\hline

\end{tabular}\end{center}

\protect\caption{Benchmark values for transport coefficients: longitudinal and transverse
diffusion, and energy gradient parameter. Boltzmann equation (BE)
results are compared with the SSMC results. Quoted error is twice
the standard error in the mean of the SSMC results. Comparisons between
BE and SSMC are shown as a third sub-column for each type of transport
coefficient.}
\end{sidewaystable*}

\end{document}